\documentstyle[12pt]{article}
\usepackage{amssymb}
\newcommand{\beq}{\begin{equation}}
\newcommand{\eeq}{\end{equation}}

\begin{document}

\section*{The torus and the Klein bottle amplitude of permutation orbifolds}
\begin{center} Zoltan Kadar (zkadar@cs.elte.hu) \end{center}
\begin{center} Eotvos University Budapest HUNGARY \end{center}
\begin{center} \small{\bf{Abstract}} \end{center}
\small{The torus and the Klein bottle amplitude coefficients are
computed in permutation orbifolds of RCFT-s in terms of the same
quantities in the original theory and the twist group. An explicit
expression is presented for the number of self conjugate primaries in
the orbifold as a polynomial of the total number of primaries and the
number of self conjugate ones in the parent theory. The formulae in
the $Z_2$ orbifold illustrate the general results.
\bigskip

Permutation orbifolds have been investigated in the last couple of years, as they are not only a special class of Rational Conformal Field Theories, 
but they are also closely related to second quantisation of strings (\cite{dmvv}). The one loop amplitude is a starting point there, its form - that is, its 
dependence on the characters of the primary fields of the corresponding Conformal Field Theory - is determined
from general principles. Finding the explicit dependence amounts to writing down the coefficients of the corresponding linear combination of the characters in
the open, the sesquilinear combination in the closed case, respectively. This is possible in a permutation orbifold - and the topic of this paper - 
in terms of the same coefficients of the "ascendant" CFT. It is useful for getting information about the structure of orbifolds and for providing 
explicit formulae which can be subject for resting conjectures about the further structure of the amplitude coefficients.      

For any RCFT ${\cal C}$ and any permutation group $\Omega<S_n$ a new CFT ${\cal C}\wr\Omega
$ can be constructed by taking the n-fold tensor product of ${\cal C}$ and identifying states
according to the orbits of the standard action of $\Omega$. The new theory is 
called the permutation orbifold of ${\cal C}$ (\cite{heid}, \cite{bori}) and every relevant quantity (e.g. conformal weights, genus one characters of the primaries,
the matrix elements of the modular transformations, the partition function etc.)
is completely determined in terms of the corresponding quantities
 of ${\cal C}$ and the twist group $\Omega$. The general case (when the twist group is nonabelian) was discussed in \cite{b1}.
  
In general, the torus amplitude of a CFT is always expressible as a sesquilinear combination
of the characters of its primaries:
\beq \label{1} Z(\tau)=\sum_{p,q}Z_{pq}\chi_p(\tau)\bar\chi_q(\tau) \eeq
where the matrix $Z_{pq}$ is invariant under the modular group and has 
 nonnegative integer elements (see eg. \cite{zepequ}).
    On the other hand, based on covering surface considerations in \cite{b1}, 
it was also shown that the partition function of the permutation orbifold ${\cal C}\wr\Omega$ of
the RCFT ${\cal C}$ ($\Omega<S_n$ is the twist group) reads
\beq \label{2} Z^\Omega(\tau)=\frac{1}{\vert\Omega\vert}\sum_{xy=yx}
\prod_{\xi\in{\cal O}(x,y)}Z(\tau_\xi). \eeq
 ${\cal O}(x,y)$ is the set of orbits (on the set: $\{1,2,...,n\}$) of the subgroup generated by 
the 2 elements $x,y$ of $\Omega$, $Z(\tau)$ is the torus 
partition function of ${\cal C}$, and $\tau_\xi$ is the modular parameter of the
covering torus corresponding to the orbit $\xi$. ($\tau_\xi=\frac{\mu_\xi\tau+\kappa_\xi}{\lambda_\xi}$
, where the three parameters corresponding to each orbit are:

$\lambda_\xi$ is the common length of the $x$ orbits in $\xi$

$\mu_\xi$ is the number of the $x$ orbits in $\xi$
 
$\kappa_\xi$ is the smallest nonnegative integer for which $y^{\mu_\xi}=x^{\kappa_\xi}$ holds in $\xi$.)  Comparing (\ref{1}) and
 (\ref{2}) gives us way to express the matrix elements $Z_{pq}$ in the orbifold as a function of those of the original CFT.  The
 primaries of the orbifold are characterized by pairs $(P,\Phi)$, where $P$ represents an orbit of $\Omega$ acting on the n tuples
 $p_1...p_n$ of primaries of ${\cal C}$, and $\Phi$ is an irreducible character of the double (\cite{bic}, \cite{dij}) ${\cal
 D}(\Omega_P)$ of the corresponding stabilizer (the subgroup of $\Omega$ which leaves P invariant).  Using the general formula of
 \cite{b1} for the genus one character of the primary field $(P,\Phi)$ \beq \chi_{(P,\Phi)}(\tau)=\frac{1}{\vert\Omega_P\vert} \sum_{x,y
 \in \Omega_P} {\bar \Phi}(x,y) \prod_{\xi \in {\cal O}(x,y)} \omega_{P(\xi)}^{-\frac{\kappa_\xi}{\lambda_\xi}} \chi_{P(\xi)}(\tau_\xi)
 \eeq we are lead to the following matrix \beq \label{4} Z_{(P,\Phi)(Q,\Psi)}=\frac{1}{\vert\Omega_P\vert\vert\Omega_Q\vert}\sum_{
 z\in\Omega\atop x,y\in\Omega_p\cap\Omega_{zQ}}\Phi(x,y){\bar \Psi}(x^z,y^z) \prod_{\xi\in{\cal O}(x,y)}Z_{P(\xi)(zQ)(\xi)} \eeq where
 $zP$ denotes the action of the group element $z$ on the n-tuple $p_1...p_n$ corresponding to $P$ (one can always work with a chosen
 representative of the orbit P, then show that the resulting formula does not depend on the choice), $P(\xi)$ is the component of P
 associated to the orbit $\xi$, and \[ \omega_p=exp(2\pi i(\Delta_p-\frac{c}{24})) \] is the exponentiated conformal weight.  The formula
 comes more or less directly from comparing the coefficients of the products of the original characters
 \[ \chi_{P(\xi1)}(\tau_{\xi1})...\chi_{P(\xi k)}(\tau_{\xi k}) \] of the same type in the two expressions.  On one hand, in eqn (\ref{1})
 the characters of the orbifold contain these products, on the other hand, writing out the bilinear expressions for each $Z(\tau_{\xi})$,
 eqn (\ref{4}) can be read off taking into account the required symmetrization, so that there should be no dependence on the actual
 representative of the orbits $P$ and $Q$.  Note that the phases appearing in the expressions of the characters in the orbifold indeed
 disappear (they are absent in (\ref{2})) as a consequence of $[Z,T]=0$ in ${\cal C}$, that is $Z_{pq}$ is nonzero only if
 $\Delta_p-\Delta_q$ is an integer, so the corresponding phases cancel each other).  The validity of the formula can be investigated via
 several checks.

 \begin{enumerate}
 \item It gives back (2) when writing down (1) for the orbifold and performing
 the summation for the pairs $(P,\Phi)$. This is a straightforward consequence of
 the method by which it was obtained. It is easily verified by using the second orthogonality
relations for the irreducible characters of the doubles ${\cal D}(\Omega_P), {\cal D}(\Omega_Q)$ 
and interchanging the sum for $P$ (and $Q$) with the one for the pairs $(x,y)$.
 \item The requirement of modular invariance.
 Since we know the explicit expression for the S and T matrices (\cite{b1}), their
 commutator with the given $Z^\Omega$ can be calculated explicitly. Using the modular
 invariance of Z in ${\cal C}$, the commutators give 0. 
 \item The important special case when $Z_{pq}$ is a permutation matrix. This
 is always the case whenever the primaries correspond to the full chiral
 algebra (\cite{nemtom}).  
 In this case the formula becomes simpler:
 \beq Z_{(P,\Phi)(Q,\Psi)} = \left \{ \begin{array}{ll} {\displaystyle \frac{1}{\vert\Omega_P\vert} \sum_{z \in N(\Omega_P)} \delta_{\Phi,\Psi^z} \prod_{i \in
 {\cal O}(\Omega_P)} Z_{P_i(zQ)_i}} & \mbox{if $\exists w \in\Omega: \Omega_P=\Omega_{wQ}$} \\ 
 0 & \mbox{otherwise} \end{array} \right. \eeq
 where $\Psi^z$ is the
 irreducible character of ${\cal D}(\Omega_P)$ defined by the relation:
 $\Psi^z(x,y)=\Psi(x^z,y^z)$, $x^z$ is an abbreviation for $z^{-1}xz$, $N(\Omega_P)$ is the normalizer of 
 $\Omega_P$ in $\Omega$ (The subgroup $\{x \in\Omega: x\Omega_P=\Omega_Px\}$) and by ${\cal O}(G)$ we mean the set of orbits of the permutation group G.
  (Note that we chose $w^{-1}Q$ in the formula
 for identifying ${\cal D}(\Omega_P)$ with ${\cal D}(\Omega_Q)$, which is possible since (\ref{4}) does not depend on the representatives of the orbits $P$ and $Q$).
 The simplification originates from the fact that any product of the form 
 $Z_{pq_1}Z_{pq_2}$ is zero whenever $q_1 \neq q_2$; it a fundamental property of permutation matrices.
 The important property of this matrix is straightforward from the formula: the
 matrix $Z_{(P,\Phi)(Q,\Psi)}$ is again a permutation matrix.   
 The map: $Z_{pq}\mapsto Z_{(P,\Phi)(Q,\Psi)}$ is a homomorphism, that is $(Z^{(1)}Z^{(2)})^\Omega
 =Z^{(1)\Omega}Z^{(2)\Omega}$ holds for the general case when Z is not necessarily
 a permutation matrix.
 It would be desirable to know the trace of the matrix for several reasons.
 (Eg. for a permutation it gives the number of its fixed points.)
 After some calculation we get
 \beq \label{7} TrZ^\Omega=\frac{1}{\vert\Omega\vert}\sum_{x,y,z\in \Omega^{(3)}}\prod_i(TrZ^i)^{i_c} \eeq   
 where $\Omega^{3}$ is the set of pairwise commuting triples form $\Omega$ and $i_c$
 is the number of cycles of length $i$ of the element $z$ on the set of orbits ${\cal O}(x,y)$. 
 \end{enumerate}
 \bigskip
 
 Although we obtained a modular invariant which satisfies several consistency
 checks, it is not unique in general, due to the linear dependence of the Virasoro
 specialized characters. Indeed, whenever we have a modular invariant $Z_{pq}$ corresponding
 to a theory, $Z_{{\bar p}q}=(S^2Z)_{pq}$, is also a good one producing the same
 torus amplitude, since the Virasoro specialized character is the same for charge conjugate fields. 
 
 What makes (\ref{4}) special is that it is unital, that is $Z_{pq}=\delta_{pq}$
 implies $Z_{(P,\Phi)(Q,\Psi)}=\delta_{(P,\Phi)(Q,\Psi)}$. This property is easily
 seen by substituting $Z_{pq}=\delta_{pq}$ into (\ref{7}). Using the notation
 $s$ for the number of primaries in ${\cal C}$ it gives 
 \beq \frac{1}{\vert\Omega\vert}\sum_{x,y,z \in \Omega^{(3)}}s^{\vert{\cal O}(x,y,z)\vert} \eeq
 which is the number of primaries in ${\cal C}\wr\Omega$ (see eg. \cite{b2}).
 
 One would like to know how $Z_{{\bar p}q}$ looks like in the orbifold. Having
 the expression for the $S$ matrix, this is not a difficult task. Introducing the notation $x^{-z}=z^{-1}x^{-1}z$, we get for the charge
 conjugation:
 \beq \label{12} S^2_{(P,\Phi)(Q,\Psi)}=\frac{1}{\vert\Omega_P\vert\vert\Omega_Q\vert}
 \sum_{z\in\Omega \atop x,y\in\Omega_p\cap\Omega_{zQ}}\Phi(x,y)\Psi(x^{-z},y^z)
 \prod_{xi\in{\cal O}(x,y)}S^2_{P(\xi)(zQ)(\xi)} \eeq
 either from the $S$ matrix or from the dimension of the space of genus 0
  holomorphic blocks for the insertion of $(P,\Phi)$ and $(Q,\Psi)$ (\cite{b2}).
  Multiplying it with (\ref{4}) we find
  \beq \label{11} Z_{\overline{(P,\Phi)}(Q,\Psi)}=\frac{1}{\vert\Omega_P\vert\vert\Omega_Q\vert}
 \sum_{z\in\Omega \atop x,y\in\Omega_p\cap\Omega_{zQ}}\Phi(x,y){\bar \Psi}(x^{-z},y^{-z})
 \prod_{\xi\in{\cal O}(x,y)}Z_{\overline{P(\xi)}(zQ)(\xi)} \eeq
 Its trace reads
\beq \label{8} Tr(S^2Z)^\Omega=\frac{1}{\vert\Omega\vert}\sum_{x,y,z\in \Omega}
\delta_{x^y,x}\delta_{x^z,x^{-1}}\delta_{y^z,y^{-1}}\prod_i(Tr(S^2Z)^i)^{i_c} \eeq
where the notations are as in (\ref{7}), and finally, when $Z_{pq}$ is a permutation
 \beq  Z_{\overline{(P,\Phi)}(Q,\Psi)}=\left \{ \begin{array}{ll} {\displaystyle \frac{1}{\vert\Omega_P\vert} \sum_{z \in N(\Omega_P)} \delta_{\Phi,\Psi^{-z}}\prod_{i \in
 {\cal O}(\Omega_P)} Z_{{\bar P}_i(zQ)_i}} & \mbox{if $\exists w \in\Omega: \Omega_P=\Omega_{wQ}$} \\
 0 & \mbox{otherwise} \end{array} \right. \eeq
  
 ($\Psi^{-z}(x,y)=\Psi((x^{-z},y^{-z})$ is the defining relation for the
 irreducible character $\Psi^{-z}$.)
 \bigskip
   
 In the case of the Klein bottle amplitude, the same procedure as comparing eqn.
 (\ref{1}) and (\ref{2}) can be done using the result of \cite{b20} for the amplitude in the orbifold:
 \beq K^\Omega(t)=\frac{1}{\vert\Omega\vert}\sum_{x,y \in \Omega}\delta_{x^y,x^{-1}}
 \prod_{\xi \in {\cal O}_{-}(x,y)}K(\frac{\lambda_{\xi}^{2}t}{\vert\xi\vert})
 \prod_{\xi \in {\cal O}_{+}(x,y)} Z(\frac{\vert\xi\vert}{2\lambda_{\xi}^{2}it}
 +\frac{\kappa_\xi}{\lambda_\xi}) \eeq  
 according to the geometric picture. ${\cal O}_{-}(x,y)$ (resp. ${\cal O}_{+}(x,y)$)
 is the set of orbits of the subgroup generated by $x$ and $y$ with odd (resp. even) 
 number of $x$ orbits. (${\cal O}_{-}(x,y)$ contains those orbits on which the commutative subgroup
 generated by $x$ and $y^2$ acts transitively, ${\cal O}_{+}(x,y)$ contains those, which fall into two orbits: $\xi+, \xi-$
 under the above subgroup). This is to be compared with the general formula
 for the amplitude in a CFT (\cite{stoeta}):
 \beq K(t)=\sum_{p}\Gamma_{p}\chi_{p}(\frac{1}{it}) \eeq
 The comparison is easier than in the case of the torus, since $K^\Omega(t)$ is a 
 linear function of the characters of the orbifold. One only has to take
 into consideration the fact that the parameter $\kappa_\xi$ is zero for the doubly covering
 torus corresponding to the orbits in ${\cal O}_{-}(x,y)$, and each element of
 ${\cal O}_{+}(x,y)$ is associated with 2 orbits in ${\cal O}(x,y^2)$ of equal length.
 All in all, we get for the coefficients in the orbifold:
 \beq \label{10} \Gamma_{(P,\Phi)}=\frac{1}{\vert\Omega_P\vert}\sum_{x,y^2 \in \Omega_P}
 \delta_{x^y,x^{-1}} \Phi(x,y^2) \prod_{\xi \in {\cal O}_{-}(x,y)}\Gamma_{P(\xi)}
  \prod_{\xi \in {\cal O}_{+}(x,y)}Z_{P(\xi+),P(\xi-)} \eeq
 (Note that the phases from the characters disappear again at ${\cal O}_{+}(x,y)$,
  due to the modular invariance of $Z_{pq}$, just like in the case of $Z^\Omega(\tau)$.)

Equations (\ref{11}) and (\ref{10}) are worth of analysing a bit further. If we insert the charge
conjugation modular invariant (that is, we substitute $Z_{pq}$ with $\delta_{pq}$ in (\ref{11})) we get the
corresponding quantity in ${\cal C}^\Omega$ (it is clear when comparing the resulting expression 
with (\ref{12})). But the trace of the charge conjugation matrix is nothing but
the number of self-conjugate primaries t (or $\sum_p\nu_p^2$, where $\nu_p$ is the 3-valued Frobenius-Schur indicator of the primary field p \cite{fs}). So we get 
\beq t^\Omega=\frac{1}{\vert\Omega\vert}\sum_{x,y,z\in \Omega}
\delta_{x^y,x}\delta_{x^z,x^{-1}}\delta_{y^z,y^{-1}}t^{\vert{\cal O}_-\vert}
s^{\vert{\cal O}_+\vert} \eeq
which is a polynomial function of the corresponding quantity in ${\cal C}$, and the total number of primaries s in 
${\cal C}$. ${\cal O}_-$ is the set of orbits generated by $x,y$ and $z$
with odd number of $<x,y>$ orbits in it ($<x,y>$ is the subgroup of $\Omega$
generated by $x$ and $y$), ${\cal O}_-$ has the same definition changing the
word "odd" to "even".

Doing the same insertion in
(\ref{10}) and substituting $\Gamma_p$ with $\nu_p$ we face $\nu_{(P,\Phi)}$ (see \cite{b20}), that is if the torus amplitude coefficients
are the Frobenius-Schur indicators of primaries in a theory, they remain the same quantities in any orbifold of it.    
This is the Orbifold Covariance Principle presented in \cite{b20}. This can be found in that article whithout the explicit formula
for $\Gamma_{(P,\Phi)}$, giving another strong argument for the Ansatz of \cite{roma} (The case was already considered in \cite{bob1}, \cite{bob12} and \cite{bob2}).
\bigskip

Let us see the explicit example of the group $Z_2$ (\cite{bori}).
There are 5 types of primary fields in the orbifold: one corresponds to the orbit $p_1p_2$ ($p_1\neq p_2$), its stabilizer is trivial, so
there is no choice for different irreducible characters but the trivial. The other four correspond to the orbit $pp$, its stabilizer is $Z_2$
itself, the double of which has four irreducible characters. Since one character is nonzero only if its first argument is in a specific
conjugacy class of the group, let 1, 2 denote those two that are nonzero only if their first argument is the unit element and 
3, 4 the other two. So summarizing the notation the primaries read:
$(p_1p_2)$, $(pp,1)$, $(pp,2)$, $(pp,3)$, $(pp,4)$
Specifying eqn (\ref{4}) for this case, we obtain
\[
\begin{array}{llll}
Z_{(p_1p_2)(q_1q_2)} & = & Z_{p_1q_1}Z_{p_2q_2}+Z_{p_1q_2}Z_{p_2q_1} & \quad \\ 
Z_{(p_1p_2)(qq,i)} & = & Z_{p_1q}Z_{p_2q} & \quad i=1,2 \\
Z_{(pp,i)(qq,i)} & = & \frac{1}{2}(Z_{pq}^2+Z_{pq}) & \quad i=1,2 \\
Z_{(pp,i)(qq,i)} & = & Z_{pq} & \quad i=3,4 \\
Z_{(pp,1)(qq,2)} & = & \frac{1}{2}(Z_{pq}^2-Z_{pq}) & \quad   
\end{array}
\]
for the independent nonzero components of the symmetric matrix. For the trace we have
\beq Tr(Z^{\Omega})=\frac{1}{2}Tr(Z)^2+\frac{1}{2}Tr(Z^2)+3Tr(Z) \eeq 
which specializes to 
\beq s^{\Omega}=\frac{1}{2}s^2+\frac{7}{2}s  \eeq
when $Z_{pq}=\delta_{pq}$ ($s$ is the total number of primaries).
There is no need to consider eqn (\ref{11}) in the case of $Z_2$, since the inverse operation is the identity. 
However the trace specializes to the following polynomial when $Z_{pq}$ is the charge conjugation
\beq t^{\Omega}=\frac{1}{2}t^2+\frac{1}{2}s+3t \eeq   
where t is the number of self conjugate primaries. This means for example that any $Z_2$ orbifold of an RCFT with no complex primary
fields has only self-conjugate ones too. 
Should one find a closed formula for $\sum_p\nu_p$ in the orbifold as well, one would have the number of real, pseudoreal and complex primaries in ${\cal C}\wr\Omega$ as 
 polynomials of the same quantities in ${\cal C}$, which is a step in classifying RCFT-s.

Finally let us enumerate the results of the present paper. Two modular invariants are found for the partition function coefficients of
a permutation orbifold expressed in terms of the original CFT. One keeps the structure $Z(\tau)=\sum_p\vert\chi_p\vert^2$, its trace therefore
concides with the number of primaries in the orbifold when $Z_{pq}=\delta_{pq}$, the other keeps the structure of 
$Z(\tau)=\sum_p\chi_p{\overline \chi_{\bar p}}$, its trace therefore is the number of self conjugate primaries in the orbifold, depending on
the same quantity and the total number of fields in the original theory. The Klein bottle coefficients are also computed and applied to
 illustrate how the OCP was used in \cite{b20}, supporting the Ansatz of \cite{roma} for the amplitude.
\section*{Acknowledgements}

I thank P. Bantay for discussions and helpful comments.


\begin{thebibliography}{16}
\bibitem{dmvv} R. Dijkgraaf, G. Moore, E. Verlinde, H. Verlinde hep-th/9608096
\bibitem{heid} A. Klemm, M. G. Schmidt, Phys. Lett. B{\bf 245} (1990) 53
\bibitem{bori} L. Borisov, M. B. Halpern, and C.Schweigert, Int. J. Mod. Phys. A{\bf 13}, 125
\bibitem{b1} P. Bantay Physics Letters B{\bf 419}, 175 (1998)
\bibitem{zepequ} P. DiFrancesco, P. Mathieu, P.Senechal, Conformal Field Theory (Springer 1997) 
\bibitem{b2} P. Bantay hep-th/9910079
\bibitem{dij} R. Dijkgraaf, V. Pasquier, P. Roche, Nucl. Phys. Proc. (Suppl.) 18B (1990) 60
\bibitem{bic} P.Bantay Phys. Lett. B{\bf 245} (1990) 477
\bibitem{nemtom} G. Moore and N. Seiberg Nucl. Phys. B{\bf 313}, 16 (1989)
\bibitem{b20} P. Bantay hep-th/0001173
\bibitem{stoeta} A. Sagnotti, in Congese `87, Non-perturbative methods in field theory, eds. G. Mack et al. (Plen, New York 1988)
\bibitem{fs} P.Bantay Phys. Lett. B{\bf 394} (1997) 87-88
\bibitem{bob1} L. R. Huiszoon, A. N. Schellekens, N. Sousa, hep-th/9909114
\bibitem{bob12} L. R. Huiszoon, A. N. Schellekens, N. Sousa, hep-th/9911229
\bibitem{bob2} T. Gannon, hep-th/9910148
\bibitem{roma} G. Pradisi, A. Sagnotti and Ya. Stanev, Physics Letters B{\bf 354}, 279 (1995)
\end{thebibliography}
\end{document}